\documentstyle[sprocl,psfig]{article}

\def\be{\begin{equation}}
\def\ee{\end{equation}}
\def\bea{\begin{eqnarray}}
\def\eea{\end{eqnarray}}

\begin{document}

\title{DO PHASE TRANSITIONS SURVIVE BINOMIAL REDUCIBILITY AND THERMAL SCALING?}

\author{L.G. MORETTO, L. PHAIR, G.J. WOZNIAK}

\address{Nuclear Science Division,\\ 
Lawrence Berkeley National Laboratory\\ 
Berkeley, CA 94720, USA}


\maketitle\abstracts{
First order phase transitions are described in terms of the microcanonical and
canonical ensemble, with special attention to finite size effects. Difficulties
in interpreting a ``caloric curve'' are discussed. A robust parameter
indicating phase coexistence (univariance) or single phase (bivariance) is
extracted for charge distributions.}
  
\section{Phase transitions, phase coexistence, and charge distributions}
Since the early studies of complex fragment emission at intermediate energies,
a ``liquid vapor phase transition'' had been claimed as an
explanation for the observed power law dependence of the fragment charge
distribution. The basis for such claims was the Fisher droplet theory
\cite{Fis67}
which was
advanced to explain/predict the clusterization of monomers in vapor. According
to this theory, the probability of a cluster of size $m$ is given by:
\begin{equation}
P(m)\propto e^\frac{-(\mu _L-\mu _V)m}{kT}m^{-\tau}e^\frac{-c_sm^{2/3}}{kT}
\end{equation}
where $\mu _L$, $\mu _V$ are the liquid 
and vapor chemical potentials, $\tau$ is
the Fisher critical exponent, $c_s$ the surface 
energy coefficient for the liquid. For $\mu _V>\mu _L$ the liquid phase is 
stable and large clusters are found. For $\mu _V<\mu _L$ the vapor is stable 
and small clusters are present. At the critical temperature the liquid-vapor 
distinction ends, $\mu _L=\mu _V$ and the surface energy coefficient 
vanishes. The cluster size distribution assumes a characteristic power law 
dependence.

A recent analysis of very 
detailed experiments has claimed not only the demonstration of a near critical 
regime, but also the determination of other critical coefficients \cite{Gil94}
besides $\tau$. 

Another recent announcement claiming the 
discovery of a first order phase transition associated with 
multifragmentation \cite{Poc95} 
has created a vast resonance. 
Because of the greater simplicity inherent to this 
subject and because 
of its relevance to some of our studies, we discuss it here in 
some detail. 

This study claims to have determined the ``caloric curve'' (sic) of a nucleus, 
namely the dependence of nuclear temperature on excitation energy. The 
temperature is determined from isotopic ratios (e.g. $^3$He$^4$He,
$^6$Li$^7$Li), while the excitation energy is determined through energy 
balance. 
The highlight of this measurement is the discovery of a 
plateau, or region of constant temperature, which is considered indicative
of a first order phase transition from liquid to vapor phase. 

Apparently, the ``paradigm'' the authors have in mind is a standard picture 
of the 
diagram temperature $T$ vs enthalpy $H$  for a one component system at 
\underline {constant 
pressure $P$}.  

It is not clear whether this experimental curve can be 
interpreted in terms of 
equilibrium thermodynamics. If this is the case, several problems arise. 
For instance, the 
claimed distinction between the initial rise (interpreted as the 
fusion-evaporation regime) and 
the plateau (hinted at as the liquid-vapor phase transition) is not tenable, 
since evaporation \underline{is} 
the liquid-vapor phase transition, and no thermodynamic difference exists 
between 
evaporation and boiling. 

Furthermore, the ``caloric curve'' requires for its interpretation an 
additional relationship 
between the variables $T$, $P$, and $V$.  More to the point, the plateau 
is a very specific feature of 
the constant pressure condition rather than being a general indicator 
of a phase transition. For 
instance, a constant-volume liquid-vapor phase transition is \underline{not} 
characterized by a plateau 
but by a monotonic rise in temperature. This can be easily proven by 
means of the Clapeyron 
equation,
together with 
the ideal-gas equation for the vapor. 

For the nearly ideal-vapor phase ($P=nT$), we write
$dP=Tdn+ndT$
where $n$ is the vapor molar density. In order to stay on the univariance 
line, we need the 
Clapeyron equation:
$dP/dT=\Delta H/(T\Delta V)$
where $\Delta H$ is the molar enthalpy of vaporization and $\Delta V$ 
is the molar change in volume from 
liquid to gas. From this we obtain:
$n\left( \Delta H-T\right)dT=T^2dn$.
At constant pressure $dn$=0, so $dT$=0. For $dn>0$, we see immediately that
$dT>0$. 
Using $dE\approx dn\Delta E$, where $\Delta E$  is the molar heat of 
vaporization at constant volume, we finally obtain:
\begin{equation}
\left.\frac{\partial T}{\partial E} \right\vert _V\approx 
\frac{T^2}{n\Delta E^2}=
\frac{1}{n\Delta S^2}.
\end{equation}
The positive value of this derivative shows that the phase 
transition at constant volume is 
characterized by a monotonic increase in temperature. 

\begin{figure}
\centerline{\psfig{file=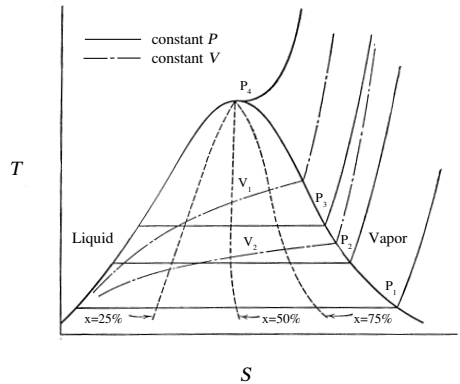,height=7.0cm}}
\caption[]{
Temperature-entropy diagram for steam. 
Curves are shown for constant pressure ($P_4>P_3>P_2>P_1$), 
constant volume ($V_1<V_2$) and 
constant percentage in 
the gas phase (dashed lines). }
\label{water}
\end{figure}

As an example, 
Fig.~\ref{water} shows a 
standard temperature $T$ vs entropy $S$ diagram for water vapor. 
The region under the bell is the 
phase coexistence region. For the constant pressure curves 
($\Delta S=\Delta H/T$), the initial rise 
along the ``liquid'' curve is 
associated with pure liquid, the plateau with the liquid-vapor phases, 
and the final rise with 
overheated vapor. The constant volume curves, however, 
($\Delta S = \Delta E/T$) 
cut across the 
coexistence region at an angle, 
without evidence for a plateau. 

Thus 
the 
reminiscence of the observed 
``caloric curve''
with ``the paradigm of a phase transition'' may be more pictorial than 
substantive, and 
indicators other than the plateau 
may be needed to 
substantiate a 
possible transition from one to two phases. More specifically, 
{\em an additional 
relationship} 
between the three variables $P,T,V$ (like $P$=const, or $V$=const, etc.)
{\em is needed to 
interpret a $T$-$E$  diagram unequivocally}.

\section{Triviality of first order phase transitions}

The great attention to the alleged discovery of first order phase transition in 
nuclei would suggest that such a phenomenon may be of great significance to 
our understanding of nuclear systems. 
In fact, it is easy to show that first order phase transitions are completely 
trivial. Here are the reasons:

1)If there are two or more phases known or even hypothetically describable, 
then there will be first order phase transitions.

2) The thermodynamics of these transitions is \underline{completely} 
determined by the thermodynamical properties of each individual isolated 
phase. These phases do not affect each other, and do not need to be in 
contact.

\begin{figure}
\centerline{\psfig{file=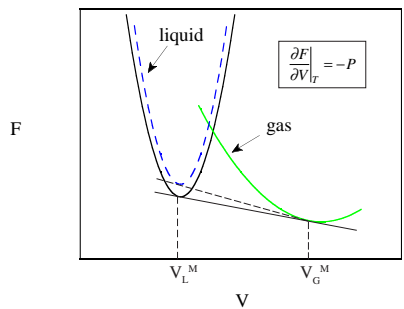,height=6.0cm}}
\vspace{-1.0cm}
\caption{The free energy as a function of molar volume for a liquid and gas.
The dashed lines refer to a drop rather than to bulk liquid.}
\label{molar}
\end{figure}

As an example, let us consider Fig.~\ref{molar}, where the molar free energy 
$F$ at constant $T$ is plotted vs the molar volume for a liquid and the gas 
phase. Stability of each phase requires each of these curves to be concave. 
In the region between 
the points of contact of the common tangent, the free energy is minimized by 
apportioning the system between the liquid and gas phase. Each phase is 
defined at the point of tangency, and the segment of the tangent between the 
two points is the actual free energy of the mixed phase. The slope of the 
common tangent is the negative of the constant pressure at which the 
transition takes place. The coexistence region is completely defined by the 
properties of the liquid at $V_L^M$ and and of the gas at $V_G^M$. 
Consequently, it is irrelevant whether 
the liquid is in contact with the vapor or not!

This discussion applies to infinite phases. However, it is simple to 
introduce finite size effects, e.g. surface effects.

%
The pressure
 of a drop is always greater than that of the infinite liquid. In
Fig.~\ref{molar}
 the common tangent (dashed) becomes steeper, in accordance with the increased
 free energy of the liquid.

\section{Microcanonical or Canonical Ensemble?}

Any good textbook of statistical mechanics contains the demonstration 
that, in the thermodynamic limit, all ensembles are equivalent, i.e. they give
 the same thermodynamic functions. 

In dealing with phase transitions in finite systems one may question whether 
this equivalence is retained. Let us review the connection between, for 
instance, the Microcanonical and the Canonical Ensemble.

Let $\rho(E)$ be the microcanonical level density. The corresponding canonical
 partition function is the Laplace transform
\begin{equation}
 Z(\beta)=\sum e^{-\beta E_i}\approx \int\rho (E)e^{-\beta E}dE
\end{equation}
The partition function is usually easier to calculate than the level density. 
However, the latter can be obtained from the former through the inverse 
Laplace Transform.
\begin{equation}
\rho (E)=\frac{1}{2\pi i}\int
\hspace{-0.125in}{\Large \circ} Z(\beta)e^{\beta E}d\beta
\label{rho}
\end{equation}

We can write Eq.~(\ref{rho}) as
\begin{equation}
S_{{\rm Micro}}=\ln\rho (E)=\ln Z_0+\beta _0E-\frac{1}{2}\ln\left( 2\pi\left.
\frac{\partial ^2\ln Z}{\partial \beta ^2}\right| _{\beta _0}\right)
\end{equation}
where $\beta _0$ corresponds to the stationary point of the integrand.
Furthermore, 
\begin{equation}
S_{{\rm Micro}}=S_{{\rm Can}}-\ln\left( 
2\pi\left.\frac{\partial ^2\ln Z_0}{\partial 
\beta ^2}\right| _{\beta _0}\right)
\end{equation}
The first term to the right is of order $N$ while the second is of order 
$\ln N$.

When $N$ goes to infinity (thermodynamic limit) one can disregard the term of 
order $\ln N$. For finite $N$ one can easily evaluate the correction term which 
turns out to be very accurate even for small $N$. For instance, consider a 
percolation system with $N$ bonds of which $n$ are broken. 
As an example of a finite system, let us take $N$=6 and $n$=3. The exact
expression yield $\rho$=20. The saddle point approximation yields $\rho$=20.6!
One can see that with little additional effort one can retain the use of the
partition function with little loss of accuracy even for the smallest systems!

Still, in the mind of physicists there is the bias that a microcanonical
approach, or its equivalent through the inverse Laplace Transform of the
Partition Function, is more correct than the canonical approach because the
former conserves energy, while the latter does not.

In fact the microcanonical distribution is given by 
\begin{equation}
P(E)=\delta\left(E(p,q)-E_0\right).
\end{equation}
The canonical distribution instead is given by 
\begin{equation}
P(E)=Ke^{-\beta E(p,q)}.
\end{equation}
In this case, there are energy fluctuations.

So, which is ultimately the ``right'' ensemble? If it does not matter, as in
the the thermodynamic limit, the point is moot. But for finite systems it
matters. However, consider the case of a small system which is a part of a
larger system. Let us call the total energy $E$ and that of the small system
$\epsilon$. Then
\begin{eqnarray}
S(E,\epsilon) & = &S(E,0)+\left.\frac{\partial 
S}{\partial\epsilon}\right|_{\epsilon=0}\epsilon+...\nonumber \\
& = & S(E,0)-\frac{\epsilon}{T}+...
\end{eqnarray}

Thus
\begin{equation}
\rho (E,\epsilon )\approx\rho (E,0)e^{-\epsilon/T}.
\end{equation}
The energy of the small system is canonically distributed, in a real, physical
sense. 
{\em The canonical, or grand 
canonical distribution very frequently has a
direct physical reality and is not an approximation to a ``more correct''
microcanonical distribution}. For instance, Na clusters 
in thermal equilibrium with a
carrier gas are canonically distributed in energy. 

What is the relevance of the
above to phase transitions? There are claims that a microcanonical approach
yields ``sharper'' phase transitions than a canonical approach, because of its
lack of energy fluctuations. However, any thermodynamic property, including
phase transitions, is defined in statistical mechanics as an ensemble average.
{\em Thus the resulting 
properties are not properties of the system alone, but they are
properties of the ensemble}. So with reference to phase transitions in
particular, arguments like ``the microcanonical ensemble yield sharper phase
transitions compared to the canonical ensemble, and because of that it is
better'' are meaningless. If the \underline{physical} ensemble is canonical,
the canonical description is the correct one, irrespective of whether it is
sharper or fuzzier than the microcanonical description.

\section{Sharpness of phases and phase transitions}

Let us consider the free energy of the liquid phase in Fig.~\ref{molar}. We can
expand about the minimum as follows:
\begin{equation}
F = F_0+\frac{1}{2}\left.\frac{\partial ^2F}{\partial
V^2}\right|_{\widehat{V}}(\Delta V)^2.
\end{equation}
The probability of volume fluctuations are then
\begin{equation}
P(V)\propto\exp\left[-\frac{(\Delta V)^2}{2\sigma_V^2}\right]
\end{equation}
where
$1/\sigma_V^2=\left.\partial ^2F/\partial
V^2\right|_{\widehat{V}}$.
Since $F\propto N$, $\sigma_V^2\propto 1/N$. Therefore important volume
(density) fluctuations are to be expected at small $N$. A cluster, or a
nucleus, 
which are not kept artificially at constant density, are going to
fluctuate substantially in density. 

At coexistence, the correlated fluctuations between the two phases make the
sharpness of the phases and of the phase transition even more washed out.

\section{A robust indicator of phase coexistence}

As we have seen, a ``generic'' caloric curve of the kind obtained in
ref.~\cite{Poc95} is of problematic interpretation because of the difficulty in
establishing the additional relation $F(V,T,P)$ associated with the evolution of
the system.

On the other hand, theoretical predictions of multifragmentation phase
transitions on the basis of calculated discontinuities in the specific heat are
suspicious because these calculations are performed at constant volume!

Furthermore, the only meaningful experimental question 
about phase transitions is
whether the system is present in a single phase or there is phase coexistence.
In thermodynamical language, we want to know whether the system is monovariant
(two phases) or bivariant (one phase).

We have found a robust indicator for just these features in the charge
distributions observed in multifragmentation. 

The charge distributions depend
both on the number $n$ of fragments in the events, and on the excitation
energy, measured through the transverse energy $E_t$ \cite{Phair95,Moretto96}.

It is possible empirically to ``reduce'' the charge distributions for $n$
fragments to that of just one, 
and to ``scale'' the transverse energy effect by means of the
following empirical equation
\begin{equation}
\sqrt{E_t}\left(\ln P_n(Z)+ncZ\right) = -F(Z)
\end{equation}
where $F(Z)$ is a universal function of $Z$, and $c$ is a
constant. This empirical equation suggests that the charge distributions can be
expressed as:
\begin{equation}
P_n(Z)\propto e^{-\frac{B(Z)}{T}-ncZ}.
\label{P_nZ}
\end{equation}
The first term in the exponent was 
interpreted \cite{Phair95,Moretto96} as an energy or enthalpy term, 
associated with the energy (enthalpy) 
needed to form a fragment. 
The second term was claimed to point to an asymptotic 
entropy associated with the combinatorial 
structure of multifragmentation. 
It was observed that a term of this form arises 
naturally in the charge distribution obtained 
by the least biased breaking of an integer $Z_0$ into $n$ 
fragments \cite{Phair95}. Such a $Z$ distribution is given approximately by: 
\begin{equation}
P(Z)=\frac{n^2}{Z_0}e^{-\frac{nZ}{Z_0}}=cn^2e^{-cnZ}.
\label{Euler}
\end{equation}
While this form obviously implies charge conservation, 
it is not necessary that charge conservation be 
implemented as suggested by Eq. (\ref{Euler}). 
In fact it is easy to envisage a regime where the 
quantity $c$ should be zero. 
Sequential thermal emission is a case in point. 
Since any fragment does not know how many 
other fragments will follow its emission, 
its charge distribution can not reflect 
the requirement of an unbiased partition of the total charge among $n$
fragments. 

On the other hand, in a simultaneous emission 
controlled by a $n$-fragment transition state \cite{Lop90}, 
fragments would be strongly aware of each other, and 
would reflect such an awareness through the charge distribution.

\begin{figure}
\centerline{\psfig{file=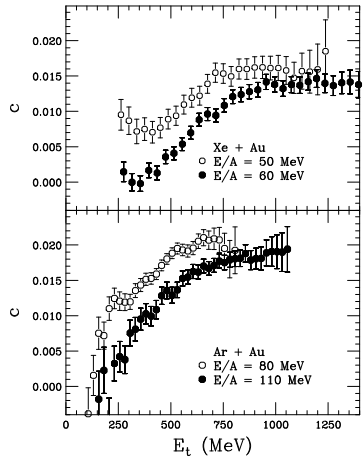,height=6.0cm,angle=180}}
\caption{Plots of the coefficient $c$ versus $E_t$ for the 
reactions $^{129}$Xe+$^{197}$Au at $E/A$=50 and 60 MeV (top panel) and
$^{36}$Ar+$^{197}$Au at $E/A$=80 and 110 MeV (bottom panel). The error bars are
statistical.
}
\label{chi2_c}
\end{figure}

The question then arises whether $c=0$ or $c>0$, 
or even better, whether one can identify a 
transition from a regime for which $c=0$ to a new regime for 
which $c>0$. 
To answer this question, 
we have studied the charge distributions 
as a function of fragment multiplicity $n$ and transverse energy $E_t$ 
for a number of systems and excitation energies. 
Specifically, we present data for the reaction $^{36}$Ar+$^{197}$Au at 
$E/A$=80 and 110 MeV and the reaction $^{129}$Xe+$^{197}$Au at 
$E/A$=50 and 60 MeV in Fig.~\ref{chi2_c}.

It is interesting to notice that for all 
reactions and bombarding energies the quantity $c$ starts at or 
near zero,  
it increases with increasing $E_t$ for small $E_t$ values, and seems to 
saturate to a constant value at large $E_t$.

This behavior can be compared to that 
of a fluid crossing from the region 
of liquid-vapor coexistence (univariant system) to the region 
of overheated and unsaturated vapor (bivariant system).
In the coexistence region, 
the properties of the saturated vapor cannot 
depend on the total mass of fluid. 
The presence of the liquid phase guarantees 
mass conservation at all average densities for 
any given temperature. 
Hence the vapor properties, and, in particular, the 
cluster size distributions cannot 
reflect the total mass or even the 
mean density of the system. In our notation, $c=0$.


On the other hand, in the region of unsaturated vapor, 
there is no liquid to insure mass conservation. 
Thus the vapor itself must take care of this conservation, 
at least grand canonically. In our notation, $c>0$. 
In other words we can associate $c=0$ with thermodynamic univariance, and $c>0$
with bivariance.

%

To test these ideas in finite systems, we have considered a finite
percolating system and a system evaporating according to a thermal binomial
scheme \cite{Moretto95,Ghetti95}. Percolation calculations \cite {Bau88} 
were performed for systems of $Z_0$=97, 160 and 
400 as a function of the percentage of bonds broken ($p_b$).
Values of $c$ were extracted 
as a function of $p_b$.

\begin{figure}
\centerline{\psfig{file=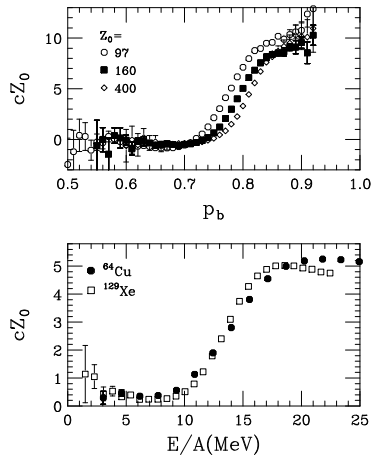,height=6.0cm,angle=90}}
\caption{top: A plot of $cZ_0$ 
versus the percentage of broken bonds $p_b$ 
from a percolation calculation \protect\cite{Bau88} for three 
systems $Z_0$=97 (circles), $Z_0$=160 (squares) and $Z_0$=400 (diamonds). 
bottom: A plot of $cZ_0$ versus excitation energy per nucleon from a binomial
evaporation calculation \protect\cite{Ghetti95} for $^{64}$Cu and $^{129}$Xe. 
The statistical 
error bars are shown for errors
larger than the symbol size.}
\label{percolation}
\end{figure}

The results are shown in Fig.~\ref{percolation}. 
For values of $p_b$ smaller than the critical (percolating) value 
($p_b^{crit}\approx $ 0.75 for an infinite system), 
we find $c=0$. 
This is the region in which a large (percolating) 
cluster is present. As $p_b$ goes above its critical 
value, the value of $c$ increases, and eventually 
saturates in a way very similar to 
that observed experimentally. 

Notice that although the phase transition in the infinite system is second
order at $p=p_c$, here the region for which $c=0$ mimics a first order phase
transition.

An evaporation calculation was also carried out for the nuclei $^{64}$Cu and
$^{129}$Xe according to the thermal binomial scheme \cite{Moretto95,Ghetti95}. 
The only
constraint introduced was to prevent at every step the emission of fragments
larger than the available source. The resulting charge distributions are very
well reproduced by Eq.(\ref{P_nZ}). The extracted quantity $cZ_0$ is plotted in
the bottom panel of 
Fig.~\ref{percolation} as a function of excitation energy per nucleon. In both
cases $cZ_0$ goes from 0 to a positive finite value (equal for both nuclei) as
the energy increases. The region where $c=0$ is readily identified with the
region where a large residue survives. On the other hand when $c>0$ there is no
surviving residue. 

These results are in striking agreement with those obtained
for percolation. For both kinds of finite systems, the univariant regime
($c=0$) is associated with the presence of a residue while the bivariant
regime ($c>0$) with the absence of a residue.


This work was supported by the Director, Office of Energy Research, 
Office of High Energy and Nuclear Physics, 
Nuclear Physics Division of the US Department of Energy, 
under contract DE-AC03-76SF00098.

\end{document}